\documentclass[longbibliography,superscriptaddress,showpacs,aps,twocolumn,prb,floatfix]{revtex4-2}

\usepackage{amssymb}
\usepackage{amsmath}
\usepackage{hyperref}
\usepackage{graphicx}
\begin{document}

\title{Thermoelectric Signatures of Kondo Physics in Geometry-Tunable Double Quantum Dots}

\author{D. Perez Daroca}
\affiliation{Gerencia de Investigaci\'on y Aplicaciones, GAIDI, CNEA, 1650 San Mart\'{\i}n, Buenos Aires, Argentina}
\affiliation{Consejo Nacional de Investigaciones Científicas y Técnicas, 1025 CABA, Argentina}

\author{P. Roura-Bas}
\affiliation{Centro At\'{o}mico Bariloche, GAIDI,  8400 Bariloche, Argentina}
\affiliation{Consejo Nacional de Investigaciones Científicas y Técnicas, 1025 CABA, Argentina}

\begin{abstract}
The equilibrium thermoelectric and spectral properties of a double quantum dot system are investigated, with the geometry continuously tuned from series to parallel via a parameter $ p $. Within the non-crossing approximation in the infinite-$ U $ limit, the Kondo peak remains robust, while satellite features and the Kondo temperature show strong sensitivity to the geometry. The Seebeck coefficient exhibits sign reversals and non-monotonic behavior as a result of the interplay between Kondo and satellite peaks. These findings underscore the role of interference and coupling asymmetry in governing transport properties, suggesting routes for geometry-based optimization in nanoscale devices.
\end{abstract}

\maketitle

\section{Introduction}

Quantum dot systems offer a highly controllable platform to explore strong correlation effects \cite{kalliakos}, quantum interference \cite{busser} and quantum simulations \cite{barthe}. In particular, double quantum dots (DQDs) have attracted considerable attention due to their ability to simulate molecular states and to exhibit a wide variety of many-body effects, such as the Kondo effect, Fano interference, and coherent electron transport \cite{tesser,donsa,Dare17,craven,sierra,dare,yada,heat,lavagna, daroca25,lomba,zhang,ghosh,mana2,daroca23,sobrino24,tosi,tetta,asym,width,wang,lombardo2025}.  The controllable nature of their coupling to external leads and to each other makes them ideal candidates for studying fundamental aspects of quantum transport, as well as for applications in thermoelectric energy conversion \cite{bene}.

The Kondo effect \cite{hewson}, arising from the screening of a localized spin by conduction electrons, plays a central role in the low-temperature physics of quantum dots. In DQD systems, the presence of two coupled dots introduces an additional energy scale associated with inter-dot tunneling, which competes with Kondo correlations and can give rise to complex spectral and transport behaviors. Furthermore, the geometry of the coupling between the dots and the leads strongly influences the system's properties, especially through interference effects that can enhance or suppress transport features.

We now turn to selected results  for DQD systems in both series and parallel configurations, where different coupling geometries give rise to a rich variety of phenomena including thermal rectification, quantum interference, and quantum phase transition.

Tesser {\it et al.} \cite{tesser} showed that heat rectification in quantum dots can be nearly perfect by tuning energy levels and cotunneling effects, while Zhang {\it et al.} \cite{zhang} found enhanced thermal rectification and negative differential conductance in parallel-coupled DQDs due to level asymmetry and strong interactions. Lavagna {\it et al.} \cite{lavagna} and Trocha {\it et al.} \cite{trocha} analyzed electrical and thermoelectric properties, highlighting the role of bonding–antibonding states and interference in enhancing thermoelectric performance and spin effects. Using numerical renormalization group, Nisikawa {\it et al.}\cite{nisikawa} revealed conductance regimes dependent on dot occupancy and coupling, complemented by Karrasch {\it et al.} \cite{karrasch} functional renormalization group approach capturing interaction-driven resonances. Li {\it et al.} \cite{li} demonstrated that small symmetry-breaking perturbations in nearly symmetric DQDs stabilize distinct current states, enabling sensitive switching. Ladrón {\it et al.} \cite{guevara} showed that destructive interference fully localizes antibonding states in symmetric parallel configurations, while Lu {\it et al.} \cite{lu} illustrated control over Fano interference and molecular state swapping via couplings and magnetic flux. Finally, Protsenko {\it et al.}\cite{protsenko} found that asymmetries in dot-lead couplings induce quantum phase transitions with distinct conductance behaviors, offering semi-analytical insight. Experimentally, double quantum dots have been implemented in various platforms, including AlGaAs/GaAs heterostructures \cite{keller}, Si-MOS devices \cite{king}, InAs/InP nanowires \cite{dorsch}, and Ge hut nanowires \cite{zhou}, among others.

In this work, we investigate the equilibrium thermoelectric and spectral properties of a DQD system where the coupling geometry is continuously tuned from a series to a parallel configuration via a control parameter $p$. Using the non-crossing approximation (NCA) in the limit of infinite on-site Coulomb repulsion, we calculate the spectral density, Kondo temperature, occupation numbers, Seebeck coefficient, and thermal conductance. Our aim is to elucidate how the interplay between geometry, inter-dot coupling, and electron correlations governs the electronic and thermoelectric behavior of the system.

\section{Model and method}\label{model} 

\begin{figure}[htbp]
\centering
\includegraphics[width=1.0\columnwidth]{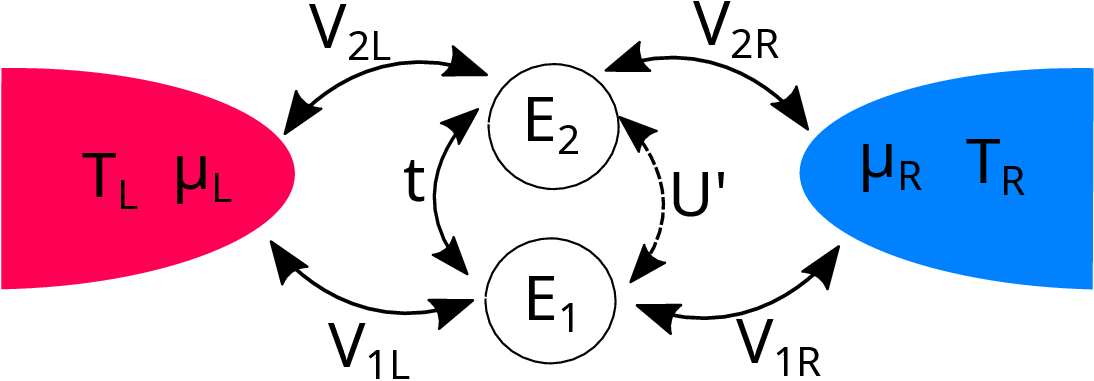}
\caption{(Color online) Schematic representation of the double quantum dot system connected in parallel to two metallic leads.} 
\label{fig1}
\end{figure}

The system displayed in Figure \ref{fig1} consists of a double quantum dot (DQD) connected in parallel to two metallic leads. The Hamiltonian describing the system is as follows

\begin{eqnarray}
H &=&H_{DQD} + H_c + H_V .  \label{ham}
\end{eqnarray}%

The first term describes the DQD, 
\begin{eqnarray}
H_{DQD}&=&\sum_{i\sigma }E_{i }d_{i\sigma }^{\dagger }d_{i\sigma }+\sum_{i }U n_{i\uparrow}n_{i\downarrow}  \\
&&+U'\sum_{\sigma\sigma' }n_{1\sigma}n_{2\sigma'}-t\sum_{\sigma}\left(d_{1\sigma }^{\dagger }d_{2\sigma }+\text{H.c.}\right)\notag\label{hdqd}
\end{eqnarray}%
where $E_{i}$ and $U$ ($U'$) are the energy levels and the inner (inter) Coulomb repulsion respectively, where 
$i =\{1,2\}$ labels the dots and $\sigma=\{\uparrow,\downarrow\}$ stands for the spin projection. 
The hopping energy between dots is represented by $t$. We assume, for simplicity, that the DQD exhibits inversion symmetry, so that $E_L = E_R \equiv E_d$.

The second term describes the conducting leads
\begin{eqnarray}
H_{c}&=&\sum_{k\nu\sigma }\varepsilon _{\nu k\sigma }\,c_{\nu k\sigma}^{\dagger }c_{\nu k\sigma },
\label{hc}
\end{eqnarray}
where $\nu =\{L,R\}$ labels the left and right leads. While the last one, represents the hybridization between each dot with its respective lead
\begin{eqnarray}
H_{V}&=&\sum_{k\nu i \sigma }\left( V_{ik\nu }\,d_{i\sigma }^{\dagger }c_{\nu k\sigma }+\text{%
H.c.}\right)  ,  \label{hv}
\end{eqnarray}%
being $V_{ik\nu }$ the tunneling energy.

The hybridization matrices that describe the interaction between the quantum dots and the metallic leads are defined by $\Gamma_{ii}^{\nu}=\pi  V^2_{i\nu}/D$ where $2D$ represents the conduction band width, the diagonal terms and $\Gamma_{ij}^{\nu}=\sqrt{\Gamma_{ii}^{\nu} \Gamma_{jj}^{\nu}}$, the off-diagonal terms.

As already stated, the central focus of this article is the transition from a parallel-coupled geometry to a series-coupled geometry. We implement this transition through a dimensionless parameter $p$, which varies continuously between 0 and 1, where $p = 0$ corresponds to the series configuration and $p = 1$ to the parallel configuration. This parameterization modifies the tunneling energies as  followed	$V_{1L}=V_{2R}=pV$ and	$V_{1R}=V_{2L}=V$.

The spectral density is a key tool for characterizing the electronic properties of the system. Here, we derive a general expression for one dot spectral density using the advanced and retarded Green's functions, 
\begin{equation}
	\rho_{ii}(\omega)=-\frac{1}{2\pi}Im[G_{ii}^>(\omega)-G_{ii}^<(\omega)],
	\label{eq:rho}
\end{equation}
and the total spectral density is $\rho=\rho_{11}+\rho_{22}$.

Another useful property of the system is the occupation number, which describes the electron population in a quantum dot,  and is given by

\begin{equation}
	n_{ii}=\frac{1}{\pi Q} \int \operatorname{Im}(G^{<}_{ii}(\omega)) \, d\omega .
	 \label{eq:occ}
\end{equation}

The Onsager matrix $\hat{L}$, which relates currents to thermodynamic forces, takes the form

\begin{equation} \label{l}
\hat{L}=-\frac{k_B T}{2 h} \int d \omega \left(\begin{array}{cc} 
e & e \omega \\
\omega & \omega^2 \end{array} \right) \tau(\omega) \; \frac{\partial f(\omega)}{\partial \omega},
\end{equation}
where the transmission function is
\begin{equation}
	\tau(\omega)= \sum_{i\nu} \Gamma_{ii}^{\nu}(\omega) \rho_{ii}(\omega)
\end{equation}
and $f(\omega)$ is the Fermi-Dirac distribution. From this matrix, we will derive the transport coefficients calculated in this work. It is important to stress that all transport coefficients presented in this work are obtained strictly within the linear response regime, i.e., for small voltage and temperature gradients.

For the calculation of  Green's functions in Equation \ref{eq:rho} and \ref{eq:occ}, we use the non-crossing approximation (NCA) \cite{win}. This method  has been successfully employed to study the  equilibrium and non-equilibrium properties of various systems \cite{hettler}. These include two-level quantum dots and C$_{60}$ molecules exhibiting a quantum phase transition \cite{sitri1,sitri2,serge}, nanoscale silicon transistors \cite{tetta}, and vibrating molecules \cite{vibra,desint19}, among others \cite{lomba,choi}. The NCA also correctly captures the scaling behavior of the conductance at low bias voltage $V$ and temperature $T$ \cite{roura10}. More recently, it has been applied to investigate three-terminal thermoelectric engines designed for energy harvesting \cite{Dare17}, as well as to control thermopower and related properties \cite{lomba}.

The main drawbacks of the Non-Crossing Approximation (NCA) in the context of the model under study emerge when considering moderate positive values of $E_d$ and finite, moderately strong Coulomb repulsion $U$. When $E_d$ is positive, the impurity self-energy develops an unphysical positive imaginary component, which leads to an artificial peak in the spectral function $\rho(\omega)$ at the Fermi level. Comparable spurious features also appear when a finite magnetic field is applied \cite{win}.
Moreover, for finite $U$, the NCA fails to capture the correct dependence of the Kondo temperature $T_K$ on the model parameters, indicating the necessity of incorporating vertex corrections.\cite{pruschke89,haule01}
However, in the present work we focus on the Kondo regime, characterized by $-E_d, E_d + U \gg \Delta$, and we take the limit $U \rightarrow \infty$, which allows us to bypass these issues. A minor limitation is that the height of the Kondo resonance tends to be overestimated by approximately $15\%$ with respect to the value predicted by the Friedel sum rule.

To analyze the system within the NCA while avoiding vertex corrections, we first diagonalize $H_{\mathrm{DQD}}$ and retain only two neighboring configurations, corresponding to the limit $U \to +\infty$ (See Appendix \ref{app_diag} and Ref \cite{daroca23}).  In addition to $U$ being very large, we also take $U'$ to be sufficiently large to prevent the occupancy of more than one particle in the double dot. As a result, we consider only fluctuations between 0 and 1 particle in the DQD.

\section{Results}

In what follows we use $\Gamma=1$ as our unity of energy,  $E_{d}=-4\Gamma$ for the energy level of the QD and $U\rightarrow \infty $. We set the Fermi energy $\epsilon_F = 0$ as the reference point for one-particle energies.  We also take $e=h=k_B=1$. The hopping parameter, $t$ and the frequency, $\omega$  have energy units. As usual, we assume a constant conduction density of states with bandwidth $2D$, with $D=10$. As we focus in equilibrium properties, we set $\Delta V=0$ and $\Delta T=0$.

\subsection{Spectral density}

In a double quantum dot system where the connection geometry transitions from parallel to series through a parameter $ p $,  the spectral density reveals a rich interplay between the Kondo effect and the formation of satellite peaks. In all cases, a central Kondo peak remains near $ \omega \approx 0 $, but the presence and characteristics of the satellite peaks depend sensitively on both the inter-dot coupling $ t $ and the geometry parameter $ p $.

\begin{figure}[htbp]
  \centering
	\includegraphics[width=0.95\columnwidth]{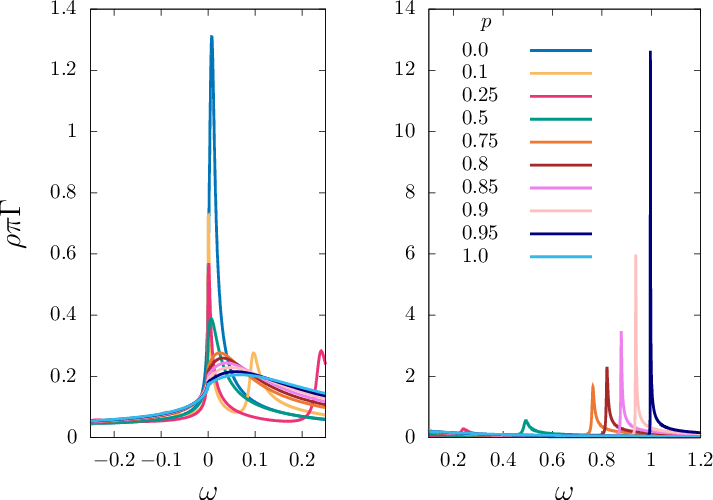}
\caption{(Color online) Spectral density for several values of $p$ in the case $t=0$.  The panels separate the Kondo resonance around $\omega\approx 0$ (left) and the satellite peaks structures (right).} 
  \label{fig:rhot0}
\end{figure}

\begin{figure}[htbp]
  \centering
    \includegraphics[width=0.95\columnwidth]{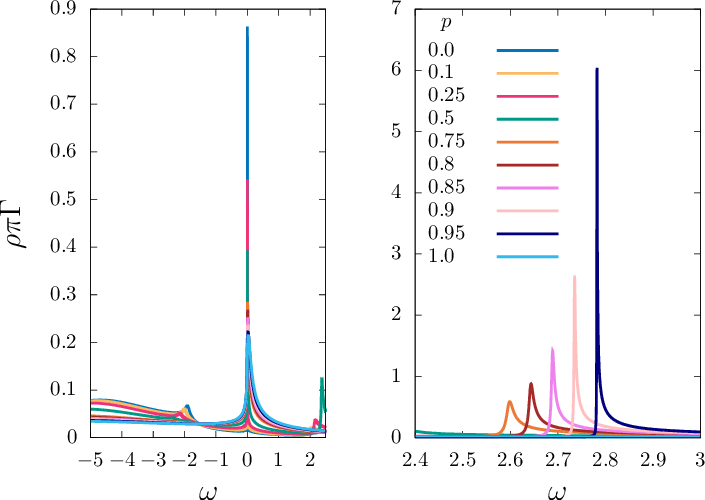}
\caption{(Color online) Spectral density for several values of $p$ with finite inter-dot coupling $t=1$. The panels separate the Kondo resonance around $\omega\approx 0$ (left) and the satellite peaks structures (right). }
  \label{fig:rhot1}
\end{figure}

Figures \ref{fig:rhot0} and \ref{fig:rhot1} display the total spectral density for different values of the geometry parameter $p$ in the cases $t=0$ and $t=1$, respectively. In both figures the panels are split to separately emphasize the Kondo resonance at $\omega \approx 0$ (left panel) and the satellite peaks (right panel). 

For $t=0$ (figure \ref{fig:rhot0}), when the dots are uncoupled, the spectrum is dominated by the Kondo peak at the Fermi level. As $p$ departs from the pure series or pure parallel limits, additional resonances appear at finite energies $\omega \simeq \pm p\Gamma$. The positive-energy satellite strengthens steadily as $p$ increases, becoming both higher and narrower up to $p \approx 0.95$, and then disappears abruptly at $p=1$. The negative-energy satellite, by contrast, remains much weaker throughout the whole range of $p$ values.

For $t=1$ (figure \ref{fig:rhot1}), the inter-dot tunneling produces bonding and antibonding molecular states, which give rise to satellites at approximately $\omega \simeq \pm(2t+p\Gamma)$. As in the $t=0$ case, the positive-energy satellite grows with $p$ and persists up to very large $p$ values, but it vanishes when $p=1$. The negative-energy satellite evolves differently: its weight increases as $p$ decreases, becoming relatively stronger near the series configuration.

The disappearance of the satellites at $p=1$ follows directly from the molecular hybridizations given in equation \ref{hy}. In the perfectly parallel configuration the  couplings of one molecular mode vanish, which means that this state is completely decoupled from the leads and its spectral signature does not appear.

\subsection{Kondo temperature}

A key energy scale characterizing strongly correlated quantum dot systems is the Kondo temperature, $T_K$. It defines the onset of many-body screening of localized spins by conduction electrons, giving rise to the formation of a narrow resonance at the Fermi level. We analyze how $T_K$ evolves as a function of $t$ for different values of $p$. This temperature was  extracted from the half-width at half-maximum (HWHM) of the Kondo peak in the spectral density.

\begin{figure}[h!]
\begin{center}
\includegraphics[width=0.95\columnwidth]{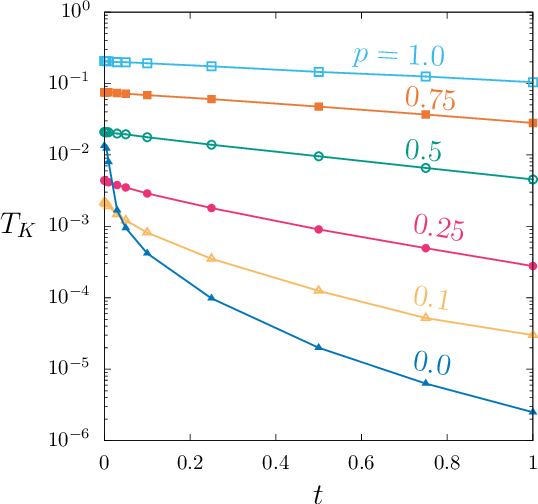}
\end{center}
	\caption{(Color online) Kondo temperature  vs $t$ for several values of $p$.}
\label{fig:TKs}
\end{figure}

Figure \ref{fig:TKs}  shows how the Kondo temperature  depends on the inter-dot coupling  for various values of the parameter $p$. For $p = 1$, corresponding to a fully symmetric configuration where both dots couple equally to the leads, the Kondo temperature remains large and nearly constant as $t$ increases, indicating that the many-body screening effect is robust and not significantly affected by the inter-dot coupling over the range shown. As $p$ decreases from 1, introducing asymmetry in the coupling between the dots and the leads, the Kondo temperature begins to drop more noticeably as $t$ increases. This behavior reflects the increasing competition between the formation of a local singlet between the two dots and the Kondo screening of the local moments by the conduction electrons. The more asymmetric the geometry (smaller $p$), the more sensitive $T_K$ becomes to the growth of $t$, and the faster it decreases as the inter-dot coupling favors the direct singlet over the screening effect. In the limit $p = 0$, which corresponds to a series-like connection where the dots are maximally asymmetric in their coupling to the leads, the Kondo temperature is low even for small $t$ and decreases rapidly as $t$ increases.

\subsection{Occupation number}

In this subsection, we examine the evolution of the occupation numbers $n_+$ and $n_-$ associated with the bonding and antibonding states, respectively, as a function of the inter-dot coupling $t$ for different values of the geometry parameter $p$. This analysis allows us to identify how the molecular character of the states changes across the transition from series to parallel configurations, and how geometry influences the degree of localization or delocalization in the system.

\begin{figure}[htbp]
  \centering
    \includegraphics[width=0.95\columnwidth]{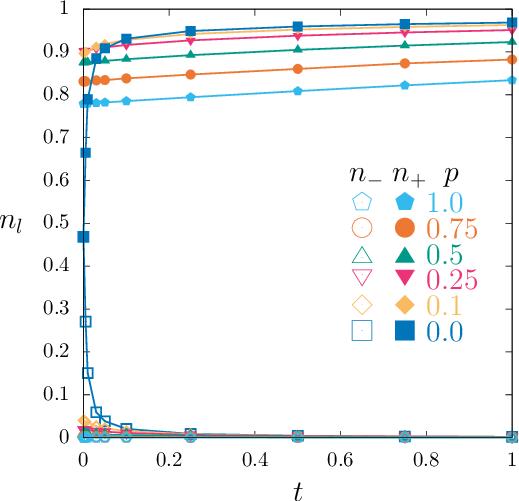}
  \caption{ Occupation number ($n_+$ and $n_-$) as a function of $t$  for various values of $p$  and $T/T_K=0.1$.}
  \label{fig:occ}
\end{figure}

Figure~\ref{fig:occ} presents  $n_-$ and $n_+$, calculated with Equation \ref{eq:occ}, as a function of $t$ for several values of $p$ and $T/T_K=0.1$. In the purely series case ($ p = 0 $), the system does not distinguish between bonding and antibonding at $ t = 0 $, and both states have identical occupation numbers. As $ t $ increases, the bonding state becomes energetically favorable, and the occupation rapidly concentrates in it, with $ n_+ $ approaching 1 and $ n_- $ approaching 0. In contrast, for $ p > 0 $, the geometry itself introduces a difference between bonding and antibonding even at $ t = 0 $, and their occupation numbers are no longer equal. As $ t $ increases, the system still tends to favor the bonding state, but the redistribution of occupation is more gradual, and the antibonding state retains a small but non-negligible occupation over the range of $ t $ shown. This residual occupation can be attributed to both the level broadening of the molecular states and temperature. This reflects the competition between molecular singlet formation driven by $ t $ and the influence of geometry on the coupling of the molecular states to the leads.

Furthermore, the occupation numbers confirm that the system remains firmly in the Kondo regime for $p = 1$. The total occupation per dot remains close to 1 ($n_{\text{total}} \approx 1$), consistent with the deep impurity level $E_d = -4\Gamma \ll 0$. The Kondo temperature $T_K$ is significantly smaller than $|E_d|$, which rules out the mixed-valence regime where charge fluctuations would lead to substantial deviations from single occupancy.

\subsection{Seebeck coefficient}

The Seebeck coefficient, also referred to as the thermopower, is a key thermoelectric quantity that measures the voltage generated in response to a temperature gradient. We investigate the behavior of the Seebeck coefficient as a function of temperature for different values of the geometry parameter $p$, focusing on two limiting cases of inter-dot coupling: $t = 0$ and $t = 1$. By analyzing its temperature dependence in relation to the underlying spectral density, we uncover how quantum interference, molecular hybridization, and geometry affect the thermoelectric response of the system.

\begin{figure}[htbp]
  \centering
    \includegraphics[width=0.95\columnwidth]{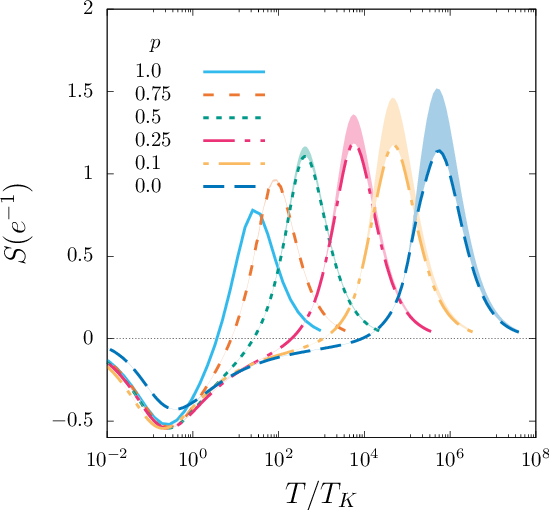}
  \caption{ Seebeck coefficient as a function of temperature   for various values of $p$  and $t=1.0$.}
  \label{fig:St1}
\end{figure}

The Seebeck coefficient as a function of temperature for $t = 1.0$ is presented in Figure~\ref{fig:St1}. This coefficient was calculated using the components of the Onsager matrix (see Equation \ref{l}) as follows:
\begin{equation}
 S = \frac{L_{12}}{T L_{11}}.
 \label{eq:S}
\end{equation}
To properly interpret this figure, it is essential to refer to the spectral density shown in Figure~\ref{fig:rhot1}. This comparison makes it possible to identify how the different features of the spectral density — the Kondo peak, the charge transfer peak, and the satellite peaks — influence the behavior of $S$ as a function of $p$.

Each curve exhibits both a positive and a negative contribution: the negative one is associated with the Kondo peak, while the positive contribution arises from the charge transfer and satellite peaks. Except for $p = 0$, all the curves collapse below $T/T_K = 1$. This behavior is directly related to the collapse of the Kondo peak in the spectral density.

The value of $ T/T_K $ at which the Seebeck coefficient changes sign shifts to the right by several orders of magnitude as $ p $ decreases. This shift reflects the evolution of $ T_K $ with $ p $, as already seen in Figure \ref{fig:TKs}. The influence of the charge-transfer peak is manifested in the width and height of the positive peaks of the Seebeck coefficient. The influence of the satellite peak in the spectral density on the positive side of $\omega$ is highlighted in the figure by the shaded areas. Specifically, the shaded region corresponds to the difference between the $S$ curves calculated with the full spectral density and those obtained using a modified spectral density $\rho_-$, which excludes the positive satellite peak. From this comparison, we observe that as $p$ decreases, the intensity of the satellite peak is reduced, and consequently, its ability to counterbalance the charge transfer peak diminishes.

\begin{figure}[htbp]
  \centering
    \includegraphics[width=0.95\columnwidth]{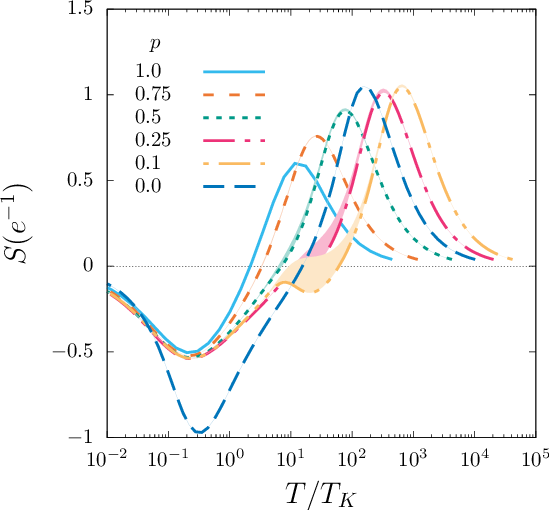}
  \caption{ Seebeck coefficient as a function of temperature   for various values of $p$  and $t=0.0$.}
  \label{fig:St0}
\end{figure}
Figure \ref{fig:St0} depicts  the Seebeck coefficient for $t = 0$ and various values of $p$. A behavior similar to the $t = 1$ case is observed for temperatures below $T/T_K = 1$, where the curves for $p > 0$ collapse. In contrast, the $p = 0$ case in this regime yields values that are lower by a factor of approximately 2.

At higher temperatures, we observe that the temperature at which the Seebeck coefficient changes sign increases as $p$ decreases. Unlike the $t = 1$ case, there is no visible contribution from the satellite peaks to the positive maxima of $S$. Instead, the effect of the satellite peaks manifests as the emergence of additional minima. This is particularly evident for $p = 0.1$.

\begin{figure}[htbp]
  \centering
    \includegraphics[width=0.95\columnwidth]{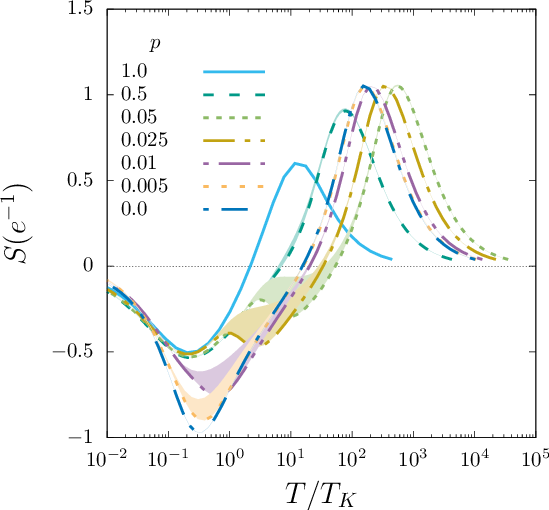}
  \caption{ Seebeck coefficient as a function of temperature   for various values of $p$, focusing on values close to $0$  and $t=0.0$.}
  \label{fig:St0chicos}
\end{figure}

This behavior is related to the fact that, in this case, the satellite peaks are located much closer to the Kondo peak, especially for small values of $p$. This effect is more clearly seen in Figure \ref{fig:St0chicos}, where we plot $S$ for even smaller values of $p$. For instance, in the cases $p = 0.05$ and $p = 0.025$, the double-minimum structure induced by the satellite peak is clearly visible.

Once again, the shaded area highlights the contribution of the positive satellite peak to the Seebeck coefficient. For smaller $p$, when the satellite peak lies closer to the Kondo peak, the effect is no longer the formation of a second minimum, but rather an enhancement (deepening) of the single minimum.

\subsection{Thermal conductance}

The thermal conductance $\kappa$ characterizes the ability of the system to transport heat in response to a temperature gradient. Unlike the Seebeck coefficient, which is sensitive to the asymmetry of the spectral density, $\kappa$ is determined by the overall spectral weight and its distribution over a wide energy range. It thus provides complementary information about the energy-dependent transmission properties of the system. The thermal conductance was calculated using the Onsager matrix components from Equation \ref{l}, yielding:

\begin{equation}
 \kappa= \frac{1}{T^2} \frac{\mbox{det} \hat{L}}{L_{11}}.
\end{equation}

\begin{figure}[htbp]
  \centering
    \includegraphics[width=0.95\columnwidth]{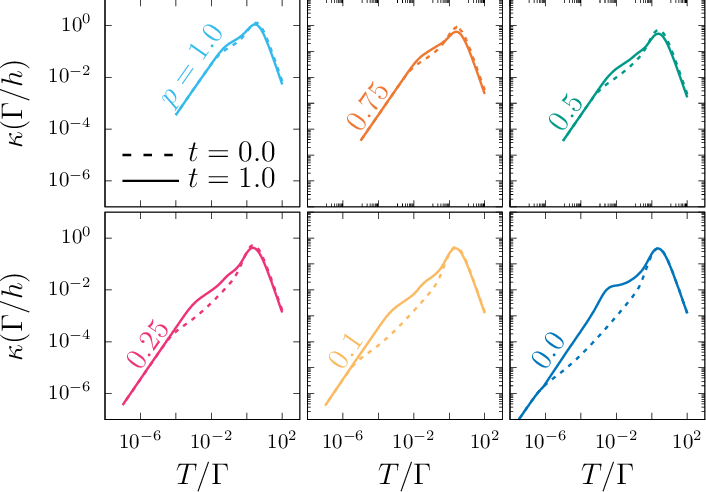}
  \caption{ Thermal conductance as a function of temperature (in units of $\Gamma$, with $k_B=1$) for various values of $p$.}
  \label{fig:kts}
\end{figure}

We analyze the thermal conductance as a function of temperature for different values of the geometry parameter $p$, considering both $t = 0$ and $t = 1$. This is displays in Figure \ref{fig:kts}. As $p$ decreases, the temperature range over which the curves for the two values of $t$ differ becomes broader. At very low temperatures, $\kappa$ is essentially the same regardless of $p$, at least for these two values of $t$.

The non–monotonic temperature dependence of the thermal conductance originates from the action of the thermal window function, $-\partial f/\partial \omega$, which selects the energy range effectively contributing to transport. At very low temperatures the window is too narrow and only probes the Kondo resonance at the Fermi level; although transmission is strong, the restricted window keeps $\kappa$ small. In this regime, all geometries yield the same result since the system behaves as a single SU(2) Kondo impurity. As the temperature increases, the window broadens and overlaps both with the Kondo peak and with the tails of high–energy spectral structures, enhancing the heat current. This interplay produces the robust maximum around $T \sim \Gamma$ observed in figure \ref{fig:kts}. For even higher temperatures, $T \gtrsim |E_d|$, the thermal window becomes too broad and the relevant spectral weight is diluted, so $\kappa$ decreases. In this regime, the system effectively behaves as a degenerate single level \cite{daroca23}.

The influence of the satellite peaks is noticeable in the curves for $t = 0$ and $p < 0.25$. This effect slightly enhances the maxima of $\kappa$, although this increase is not shown explicitly in the figure, as in the case of the Seebeck coefficient, for the sake of clarity.

\section{Conclusions}

We have studied the equilibrium thermoelectric and spectral properties of a double quantum dot system undergoing a continuous transition from parallel to series coupling geometry, controlled by a tunable parameter $p$. Using the non-crossing approximation in the limit of infinite on-site Coulomb repulsion, we explored the interplay between Kondo correlations, inter-dot coupling $t$, and geometric asymmetry.

Our analysis shows that the spectral features of a double quantum dot system are highly sensitive to both the inter-dot tunneling and the connection geometry. The central Kondo resonance remains robust near $\omega \approx 0$ across all configurations, while the finite-energy satellite peaks exhibit a nontrivial dependence on the geometry parameter $p$ and the inter-dot coupling $t$. In particular, the positive-energy satellites increase in intensity and sharpen as $p$ grows, but vanish abruptly in the purely parallel configuration ($p=1$), whereas the negative-energy satellites behave differently depending on $t$ and $p$, becoming more pronounced near the series geometry. This behavior is fully consistent with the underlying molecular hybridizations: in the parallel limit, certain molecular modes decouple from the leads, causing their spectral signatures to disappear. These findings highlight the intricate interplay between Kondo physics and molecular orbital formation in tunable double quantum dot setups.

We find that the Kondo temperature $T_K$ remains high and nearly constant for symmetric configurations ($p = 1$) but is strongly suppressed for small $p$ as the competition with inter-dot singlet formation becomes dominant. The occupation of molecular states further reflects this transition, with the bonding state becoming increasingly populated as $t$ grows, especially in asymmetric setups.

Thermoelectric response, in particular the Seebeck coefficient, is shown to be highly sensitive to spectral asymmetries. At low temperatures ($T < T_K$), all curves collapse for $p > 0$, while at higher temperatures, the satellite and charge-transfer peaks induce characteristic sign changes and nontrivial temperature dependence. The contribution of the positive-energy satellite peak was explicitly isolated, confirming its essential role in shaping the thermopower at intermediate temperatures.

The thermal conductance $\kappa$ reflects the total spectral weight over a broad energy range rather than spectral asymmetry. It exhibits a non-monotonic temperature dependence with a maximum around $T \sim \Gamma$, originating from the interplay between the thermal window, the Kondo resonance, and higher-energy spectral features. At very low temperatures, $\kappa$ is largely insensitive to the geometry parameter (p), since the system effectively reduces to a single SU(2) Kondo impurity. As the temperature increases, differences between $t=0$ and $t=1$ become more pronounced, and satellite peaks can contribute to enhancing the conductance. At even higher temperatures, the system crosses over to the regime of an effectively degenerate single level, where $\kappa$ decreases.

In summary, our analysis highlights the rich interplay between quantum interference, strong correlations, and geometry in nanoscale thermoelectric devices. The ability to control the coupling configuration offers a pathway to control thermoelectric efficiency and spectral features via geometric engineering. Future research could explore non-equilibrium effects, the role of charge fluctuations between one and two particles, and the impact of asymmetries in the dot energy levels (i.e., $E_1 \ne E_2$), which are expected to further enrich the transport behavior and open new avenues for functional design. Moreover, although we have not presented explicit results for the figure of merit $ZT$, a systematic study of this quantity would be a natural extension of the present work. Finally, benchmarking our predictions against numerically exact NRG calculations would provide a valuable quantitative validation and is left for future investigations.

\appendix

\section{Diagonalization of $H_{DQD}$}
\label{app_diag}
In this appendix, we analyze the isolated double quantum dot (DQD) system. We begin by restricting to the fluctuations between states with zero and one particle. The Hamiltonian describing the one-particle sector of the DQD system, denoted as $H_{DQD}^1$, is given by

\begin{equation}
	H_{DQD}^1 = (d_{1\sigma}^\dagger, d_{2\sigma}^\dagger) A \begin{pmatrix} d_{1\sigma} \\ d_{2\sigma} \end{pmatrix},
\end{equation}

where the matrix $A$ is defined as
\begin{equation}
	A =  \begin{pmatrix} E_d & -t \\ -t & E_d \end{pmatrix}.
\end{equation}

After diagonalizing the Hamiltonian, we obtain:

\begin{equation}
	H_{DQD}^1 = (\xi_{-\sigma}^\dagger, \xi_{+\sigma}^\dagger) \begin{pmatrix} E_d-t & 0\\ 0 & E_d+t \end{pmatrix} \begin{pmatrix} \xi_{-\sigma} \\ \xi_{+\sigma} \end{pmatrix}
\end{equation}

\begin{figure}[htbp]
\begin{center}
\includegraphics[clip,width=9cm]{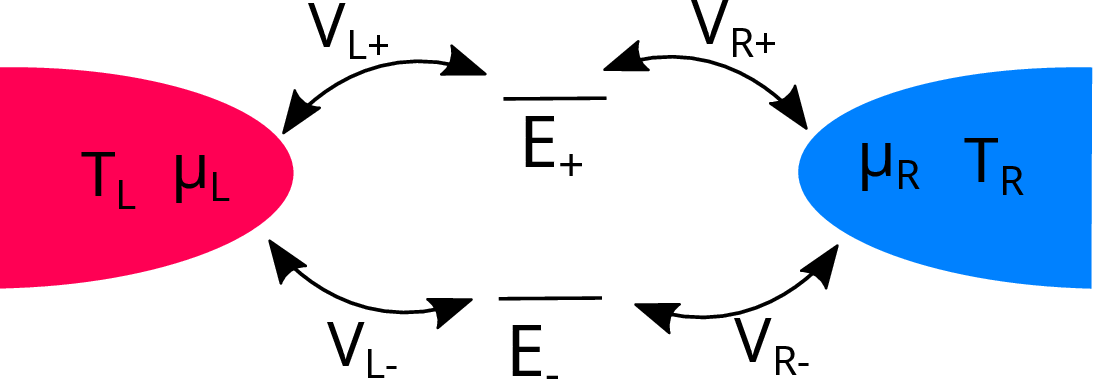}
\end{center}
\caption{(Color online) Schematic representation of the equivalent system.} 
\label{fig2}
\end{figure}

 The original operators $d_{1\sigma}^{\dagger}$ and $d_{2\sigma}^{\dagger}$ can now be expressed in terms of $\xi_{\pm \sigma}^{\dagger}$ as

\begin{equation}
	d_{1\sigma}^{\dagger}=\frac{1}{\sqrt{2}}(\xi_{-\sigma}^{\dagger}-\xi_{+\sigma}^{\dagger})
\end{equation}
\begin{equation}
	d_{2\sigma}^{\dagger}=\frac{1}{\sqrt{2}}(\xi_{-\sigma}^{\dagger}+\xi_{+\sigma}^{\dagger}).
\end{equation}

In terms of these new operators, the one-particle sector of the Hamiltonian is rewritten as:
\begin{equation}
\sum_{\xi\sigma }E_{\xi}n_{\xi\sigma }
\end{equation}
where $E_{\pm}=E_d\pm t$.

The full Hamiltonian for the one-particle sector (see schematic representation in Figure \ref{fig2}) is given by
\begin{equation}
	H_{\xi}^1=\sum_{\xi\sigma }E_{\xi}n_{\xi\sigma } + H_c +H_V^{\xi}
\end{equation}

whit the tunneling term written as
\begin{equation}
	H_V^{\xi}=\sum_{k\nu i \sigma } \sum_{m=\pm} \left( V_{ik\nu }a_{im} \,\xi_{m\sigma }^{\dagger }c_{\nu k\sigma }+\text{H.c.}\right) 	
\end{equation}

The coefficients $a_{im}$ are then given by
\begin{align}
	a_{1-}=\frac{1}{\sqrt{2}}  &     &a_{2-}=\frac{1}{\sqrt{2}} \nonumber\\
	a_{1+}=-\frac{1}{\sqrt{2}} &     &a_{2+}=\frac{1}{\sqrt{2}}
	\label{hy}
\end{align}

The hybridization coefficients $V_{\nu m}=V_{1\nu}a_{1m}+V_{2\nu}a_{2m}$ with the incorporation of the $p$ parameter  reduce to
\begin{align}
	V_{L-} &= \frac{1}{\sqrt{2}}(pV_{1L}+V_{2L}), & V_{L+} &= \frac{1}{\sqrt{2}}(-pV_{1L}+V_{2L}), \nonumber \\
	V_{R-} &= \frac{1}{\sqrt{2}}(V_{1R}+pV_{2R}), & V_{R+} &= \frac{1}{\sqrt{2}}(-V_{1R}+pV_{2R}).
	\label{hybcoeE1E2}
\end{align}

\bibliography{ref}

\end{document}